\documentclass[aps,preprint,tightenlines,groupedaddress,nofootinbib,byrevtex,showpacs]{revtex4}
\usepackage{graphicx,comment}

\usepackage{amssymb,latexsym}
\usepackage{amsmath,amsbsy}

\begin{document}

\unitlength=1mm

\def\a{{\alpha}}
\def\b{{\beta}}
\def\d{{\delta}}
\def\D{{\Delta}}
\def\e{{\epsilon}}
\def\g{{\gamma}}
\def\G{{\Gamma}}
\def\k{{\kappa}}
\def\l{{\lambda}}
\def\L{{\Lambda}}
\def\m{{\mu}}
\def\n{{\nu}}
\def\o{{\omega}}
\def\O{{\Omega}}
\def\S{{\Sigma}}
\def\s{{\sigma}}
\def\th{{\theta}}

\def\ol#1{{\overline{#1}}}

\def\Dslash{D\hskip-0.65em /}
\def\dslash{{\partial\hskip-0.5em /}}
\def\vslash{{\rlap \slash v}}

\def\CPT{{$\chi$PT}}
\def\QCPT{{Q$\chi$PT}}
\def\PQCPT{{PQ$\chi$PT}}
\def\tr{\text{tr}}
\def\str{\text{str}}
\def\diag{\text{diag}}
\def\order{{\mathcal O}}
\def\vit{{\it v}}
\def\vD{\vit\cdot D}
\def\am{\alpha_M}
\def\bm{\beta_M}
\def\gm{\gamma_M}
\def\smb{\sigma_M}
\def\smt{\overline{\sigma}_M}
\def\tb{{\tilde b}}

\def\cC{{\mathcal C}}
\def\cB{{\mathcal B}}
\def\cT{{\mathcal T}}
\def\cQ{{\mathcal Q}}
\def\cL{{\mathcal L}}
\def\cO{{\mathcal O}}
\def\cA{{\mathcal A}}
\def\cH{{\mathcal H}}
\def\cF{{\mathcal F}}
\def\cG{{\mathcal G}}
\def\cE{{\mathcal E}}
\def\cJ{{\mathcal J}}
\def\cK{{\mathcal K}}
\def\cM{{\mathcal{M}_+}}

\def\Bbar{\overline{B}}
\def\Tbar{\overline{T}}
\def\cBbar{\overline{\cal B}}
\def\cTbar{\overline{\cal T}}
\def\cA{\mathcal A}

\def\eqref#1{{(\ref{#1})}}

\preprint{NT@UW 04-019}

\title{Decuplet Baryon Masses in Partially Quenched Chiral Perturbation
  Theory}
\author{Brian C.~Tiburzi}
\email[]{bctiburz@u.washington.edu}
\author{Andr\'e Walker-Loud}
\email[]{walkloud@u.washington.edu}
\affiliation{Department of Physics\\ 
University of Washington\\
P.O.~Box 351560\\
Seattle, WA 98195-1560}

\date{\today}

\begin{abstract} 
The masses of the spin-$\frac{3}{2}$ baryons are calculated to
next-to-next-to-leading order in heavy baryon chiral perturbation
theory and partially quenched heavy baryon chiral perturbation theory.
The calculations are performed for three light flavors in the isospin limit. 
These results are necessary for extrapolating QCD and partially quenched 
QCD lattice calculations of the decuplet baryon masses.

\end{abstract}

\pacs{12.38.Gc}
\maketitle

\section{Introduction}

The study of low-energy QCD provides insight into the non-perturbative
dynamics of quarks and gluons. One of the most fundamental observables 
is the spectrum of the theory. Quarks and gluons are confined into color 
neutral hadronic states that are seen in nature. Understanding the mass spectrum 
of these lowest-lying states directly from QCD remains an on-going challenge.

A model independent tool to study low-energy QCD is through chiral 
perturbation theory (\CPT). This is an effective theory written in 
terms of low-energy degrees of freedom, e.g. the octet mesons in $SU(3)$
flavor are assumed to be the pseudo-Goldstone bosons that appear
from chiral symmetry breaking. In this effective theory observable quantities
receive both long-range and short-range contributions. The long-range contributions
arise from non-analytic meson loops, while the short-range physics is encoded in 
a number of low-energy constants (LECs). Symmetry constrains the number 
of such constants but their values must be determined from experiment
or from theoretical calculations.

Lattice QCD can provide first principles numerical determination of
QCD observables, in particular the mass spectrum of baryons. 
There have been numerous calculations of baryon masses in quenched 
QCD (QQCD) and a few in 
QCD~\cite{Aoki:1999yr, AliKhan:2001tx, Allton:2001sk, Aoki:2002uc, Zanotti:2003fx}
Partially quenched QCD (PQQCD) calculations in the baryonic sector are so far very limited. 
A problem that haunts lattice calculations is that computing power
currently restricts the simulations to quark masses that are larger
than the physical light quark masses. Therefore, to make physical predictions
it is necessary to extrapolate from the heavier lattice quark masses 
to those found in nature. For QQCD, where the fermionic determinant that
arises from the path integral is set to a constant, quenched chiral perturbation
theory (\QCPT) has been developed to aid in the 
extrapolation~\cite{Morel:1987xk,Sharpe:1992ft,Bernard:1992ep,
Bernard:1992mk,Golterman:1994mk,Sharpe:1996qp,Labrenz:1996jy}.
There is no general connection of QQCD observables to QCD because 
QQCD does not have an axial anomaly. This feature of the quenched approximation
leads to new operators in \QCPT\ that have no analogue in \CPT. Moreover
the LECs of \QCPT\ are numerically different than in \CPT. 

The problems of the quenched approximation can be remedied by using partially quenched QCD (PQQCD). 
In PQQCD contributions from sea quarks are retained and the fermionic determinant is hence no longer a constant. 
Additionally the masses of sea quarks are varied independently of the valence quarks. 
By efficaciously giving the sea quarks
larger masses it is much less costly to calculate observables than in ordinary QCD. 
The low-energy effective theory of PQQCD is partially quenched \CPT\ (\PQCPT)~%
\cite{Bernard:1994sv,Sharpe:1997by,Golterman:1998st,Sharpe:1999kj,
Sharpe:2000bn,Sharpe:2000bc,Sharpe:2001fh,Shoresh:2001ha,Sharpe:2003vy}. 
Since PQQCD retains an axial anomaly, the singlet field can be integrated out. 
Therefore the LECs of \CPT\ appear in \PQCPT.
By fitting \PQCPT\ to PQQCD lattice data one can determine the LECs
and thereby make physical predictions for QCD. 
There has been much activity recently in calculating baryon properties
in \PQCPT\ \cite{Chen:2001yi,Beane:2002vq, Chen:2002bz, Beane:2002np, Leinweber:2002qb,
  Arndt:2003vx, Beane:2003xv,Arndt:2003ww, Arndt:2003we, Arndt:2003vd,Arndt:2004as,Walker-Loud:2004hf}. 
There has also been interest in using chiral
effective theories to extrapolate quenched and unquenched lattice data on the baryon masses, 
see for example~\cite{Young:2002cj, AliKhan:2003cu, Beane:2004ks}.

In this work we calculate the masses of the decuplet baryons to
next-to-next-to leading order (NNLO) in $\chi$PT.
We also provide the NNLO
calculation of the decuplet baryon masses in PQ$\chi$PT. 
These calculations are performed in the isospin limit of $SU(3)$ and
in the partially quenched calculation we use three valence, three ghost
and three sea quarks, with two of the sea quarks degenerate.
The expressions we derive can be used to extrapolate lattice 
QCD and partially quenched lattice QCD data to the physical quark masses.
Additionally the LECs can be determined up to $\cO(m_q^2)$, where
$m_q$ represents a quark mass.

This paper has the following organization. First in Section \ref{intro},
we review the basics of \CPT\ including the inclusion of the octet and decuplet 
baryons to leading and next-to-leading order in the heavy baryon expansion.
In this work we use rank-three flavor tensors for both the octet and decuplet 
baryons. Next in Section \ref{CPT}, we calculate the decuplet masses to 
next-to-next-to-leading order in \CPT. We work in the $SU(3)$ flavor group 
and in the isospin limit. In Section \ref{PQintro}, we review \PQCPT\ in a 
way that parallels our discussion of \CPT, then in Section \ref{PQCPT}
we calculate the decuplet masses at next-to-next-to-leading order in 
\PQCPT. Finally a summary (Section \ref{summy}) ends the paper.

\section{Heavy baryon chiral perturbation theory} \label{intro}

In this section, we briefly review chiral perturbation theory. We start 
first in the meson sector and then review heavy baryon chiral perturbation 
theory to leading and next-to-leading order in the inverse baryon mass.
For the baryons, we differ from the standard formulation by embedding 
the octet states into rank-three flavor tensors.

\subsection{Pseudo-Goldstone bosons}
For massless quarks, the QCD Lagrangian exhibits a chiral symmetry,
$SU(3)_L \otimes SU(3)_R \otimes U(1)_V$, which is 
spontaneously broken down to $SU(3)_V \otimes U(1)_V$.  Chiral
perturbation theory, the low-energy effective theory of QCD, emerges by
expanding about the physical vacuum state.  
Without the explicit breaking of chiral symmetry provided by the 
quark mass term of the Lagrangian, eight particles (the pions, kaons and eta) 
would emerge as the Goldstone bosons of the broken $SU(3)_A$ symmetry.  
Given that the quark masses are small compared to the scale of chiral symmetry breaking, 
however, the lowest lying mesons emerge as an octet of pseudo-Goldstone bosons.

The pseudo-Goldstone bosons are collected in an exponential matrix
\begin{equation}
  \S = 
    {\rm exp} \left( \frac{2\ i\ \Phi}{f} \right) = \xi ^2 \ ,\quad  
      \Phi  = 
        \begin{pmatrix}
          \frac{1}{\sqrt{2}}\pi^0 + \frac{1}{\sqrt{6}}\eta & \pi^+ & K^+\\
            \pi^- & -\frac{1}{\sqrt{2}}\pi^0 + \frac{1}{\sqrt{6}}\eta & K^0\\
            K^- & \ol{K}{}^0 & -\frac{2}{\sqrt{6}}\eta
        \end{pmatrix}.
\label{eq:Sigma}
\end{equation}
With the above convention, the pion decay constant $f$ is $132$~MeV.
The effective Lagrangian describing the dynamics of these mesons at leading order in $\chi$PT is \cite{Gasser:1985gg}
\begin{eqnarray}
  \mathfrak{L}  =\ 
    \frac{f^2}{8} \tr \left( \partial^\mu \S ^\dagger
      \partial_\mu \S \right)
     + \l \, \tr \left(m_Q^\dagger  \S  + m_Q \S^\dagger \right).
\label{eqn:Lmeson}
\end{eqnarray}
Above, leading order in the power counting is $\cO(m_Q)$ and hence $\cO(m_Q) \sim \cO(q^2)$, 
where $q$ is an external pion momentum. 
In the isospin limit, the quark mass matrix $m_Q$ is given by
\begin{equation}
m_Q =  \diag (m_u, m_u, m_s).
\end{equation}
Expanding the Lagrangian Eq.~\eqref{eqn:Lmeson} to leading order,
one finds that mesons with quark content $QQ'$ are canonically normalized
when the meson masses are given by
\begin{equation}
m^2_{QQ'} = \frac{4 \lambda}{f^2} (m_Q + m_{Q'}) 
\label{eq:mesonmass}.
\end{equation}

\subsection{Baryons}

Besides the mesons, there are also three quark states in the spectrum of QCD.
To include these octet and decuplet baryons systematically into 
chiral perturbation theory, we use heavy baryon $\chi$PT (HB$\chi$PT)~%
\cite{Jenkins:1991ne,Jenkins:1991jv,Jenkins:1991es, Jenkins:1992ts}. 
For the octet, baryon fields $B(x)$ are redefined in terms of velocity dependent
fields $B_v(x)$,
\begin{equation}
B_v(x) = \frac{1+\vslash}{2} e^{i M_B \vit \cdot x} B(x)
\label{eq:Bv}\, ,
\end{equation}
where $v_\mu$ is the four-velocity of the baryon, $B$.  This field
redefinition corresponds to parameterizing the baryon 
momentum as
\begin{equation}
  p_\mu = M_B v_\mu + k_\mu,
\label{eq:Bp}
\end{equation}
where $k_\mu$ is the residual momentum.  This efficacious redefinition
eliminates the Dirac mass term for baryons
\begin{equation}
  \Bbar \left(  i\dslash - M_B  \right) B
  = \Bbar_v \, i\dslash  B_v 
  + \cO \left(\frac{1}{M_B}\right)  .
\label{eq:BvmB}
\end{equation}
From Eq.~(\ref{eq:Bv}), it is easy to verify that derivatives acting
on $B_v$ bring down powers of the residual momentum $k$.  Thus, higher
dimension operators of the heavy baryon field 
$B_v$ are suppressed by powers of $M_B$  and a consistent derivative expansion emerges.  
Henceforth we shall omit the subscript $v$ from all baryon fields treating them implicitly as
heavy baryons.

In this work we embed all baryons in rank-3 flavor tensors.  
The convenience of this choice will become readily apparent 
when we generalize to PQ$\chi$PT. 
The $SU(3)$ matrix of the lowest-lying 
spin-$\frac{1}{2}$ baryon fields is
\begin{equation}
  {\mathbf B} = 
    \begin{pmatrix}
      \frac{1}{\sqrt{6}}\Lambda + \frac{1}{\sqrt{2}}\Sigma^0 &
        \Sigma^+ & p\\
      \Sigma^- & \frac{1}{\sqrt{6}}\Lambda -
        \frac{1}{\sqrt{2}}\Sigma^0 & n\\
      \Xi^- & \Xi^0 & -\frac{2}{\sqrt{6}}\Lambda\\
    \end{pmatrix}\, .
\label{eq:B2index}
\end{equation}
These baryon states are embedded in the tensor $B_{ijk}$ in the following way~\cite{Labrenz:1996jy}
\begin{equation}
  B_{ijk}  = 
    \frac{1}{\sqrt6}\left( \varepsilon_{ijl} {\mathbf
      B}_{k}^{\hskip 0.30em l} 
    + \varepsilon_{ikl} {\mathbf B}_{j}^{\hskip 0.30em l} \right) 
\label{eq:3indexB}
.\end{equation}
The flavor tensor has the symmetry properties
\begin{equation}
     B_{ijk}  = B_{ikj} \quad {\rm and} \quad  B_{ijk}+B_{jik}+B_{kji}=0.
\end{equation}

When the spin-$\frac{3}{2}$ decuplet baryons $T$ are included
in the theory, an additional mass parameter $\D$ appears.
This parameter is the leading-order mass splitting between the octet and the decuplet in the chiral limit and
must be included in the power
counting.  We treat $\D$ as $\cO(q)$, where $q$ is a typical small pion momentum.  
The spin-$\frac{3}{2}$ decuplet baryons can be described by a
Rarita-Schwinger field, $(T^\mu)_{ijk}$, which is totally symmetric
under the interchange of flavor indices. We employ the normalization convention in which $T^{111} = \Delta^{++}$. 
The heavy baryon Rarita-Schwinger field satisfies the constraints, $v \cdot T = S \cdot T =0$, 
where $S^\mu$ is the covariant spin-vector.

The Lagrangian to leading order in the $1/M_B$ expansion
can be written as 
\begin{eqnarray}
  \mathfrak{L} &=&
    \left( \Bbar \, i\vD \, B \right) 
    + 2\am \left( \Bbar B \cM \right) 
    + 2\bm \left( \Bbar \cM B \right) 
    + 2\smb \left( \Bbar B \right) \tr(\cM)
\nonumber\\
 && -
    \left(\Tbar{}^{\mu}\left[\, i\vD - \Delta \right] T_\mu \right) 
    + 2\gm \left(\Tbar{}^\mu \cM T_\mu \right) 
    - 2\smt \left( \Tbar{}^\mu T_\mu \right) \tr(\cM)
\nonumber\\
 && +
    2\a\left(\Bbar S^\mu B A_\mu \right)\ 
    +\ 2\beta \left(\Bbar S^\mu A_\mu B \right)\ 
    +\ 2{\cal H} \left( \Tbar {}^\nu S^\mu A_\mu T_\nu \right)
\nonumber\\
 && +
    \sqrt{\frac{3}{2}}{\cC} \left[ \left( \Tbar{}^\nu A_\nu B
      \right)\ 
    +\ \left( \Bbar A_\nu T^\nu \right) \right]
\label{eq:leadlag}.
\end{eqnarray}
Above, $D_\mu$ is the chiral-covariant derivative which acts
on the $B$ and $T$ fields as
\begin{equation}
  \left( D^\mu B \right)_{ijk} = \partial^\mu B_{ijk} 
    + (V^\mu)_i^{\hskip 0.30em l} B_{ljk} 
    + (V^\mu)_j^{\hskip 0.30em l} B_{ilk} 
    + (V^\mu)_k^{\hskip 0.30em l} B_{ijl} .
\end{equation}
The vector and axial-vector meson fields appearing in the Lagrangian
are given by
\begin{equation}
  V_\mu =
    \frac{1}{2} ( \xi\partial_\mu\xi^\dagger 
      + \xi^\dagger\partial_\mu\xi ), \quad  
  A_\mu  =
    \frac{i}{2} ( \xi\partial_\mu\xi^\dagger 
      - \xi^\dagger\partial_\mu\xi )\, ,
\end{equation}
and
\begin{equation}
  \cM  = \frac{1}{2}\left( \xi^\dagger m_Q \xi^\dagger + \xi m_Q
  \xi \right)\, .
\end{equation}
In Eq.~(\ref{eq:leadlag}) the brackets $(~)$ denote a contraction
of the flavor indices and are defined in~\cite{Labrenz:1996jy}.
Such contractions ensure the proper transformations of the field
bilinears under chiral transformations.  To compare with the
coefficients used in the standard two-index baryon formulation
\cite{Jenkins:1991ne,Jenkins:1991jv,Jenkins:1991es, Jenkins:1992ts}, 
it is straightforward to show 
\begin{equation}
\a = \frac{2}{3} D + 2 F \, , \qquad \b = - \frac{5}{3} D + F \, , 
\end{equation}
and
\begin{equation}
\a_M = \frac{2}{3} b_D + 2 b_F \, , \qquad \b_M = - \frac{5}{3} b_D + b_F \, , \qquad \s_M = b_D - b_F + \s \, . 
\end{equation}

\subsection{Higher dimensional operators}
 
The Lagrangian in Eq.~(\ref{eq:leadlag}) contains some, but not all,
terms of $\cO (q^2)$.  To calculate the decuplet masses to $\cO (m_Q^2)$
we must include all $\cO (q^2)$ relevant operators, which contribute to
the mass via loops, and all $\cO(q^4)$ relevant operators which contribute 
at tree level.  The Lagrangian also includes operators of $\cO (q^3)$, but they do not
contribute to the self-energy.

The baryon mass is treated as $M_B \sim \L_\chi$.  As the
LECs are {\it a priori} unknown, we can combine
the $1/M_B$ and $1/\L_\chi$ expansions into one expansion in powers of
$1/\L_\chi$.  There is one exception: constraints from
reparameterization invariance (RI) determine the coefficients of some of the higher
dimension operators arising in the $1/M_B$
expansion~\cite{Luke:1992cs,Manohar:2000dt}.  Thus these $1/M_B$ 
corrections must be kept distinct to insure the Lorentz invariance of
the heavy baryon theory to a given order.

The heavy baryon momentum parameterization in terms of $v^\mu$ and $k^\mu$ in Eq.~\eqref{eq:Bp} 
is unique only up to $1/M_B$ corrections.  When the velocity and
residual momentum are simultaneously transformed in the following way
\begin{eqnarray}
  v \rightarrow v + \frac{\e}{M_B}\ \ ,\ \ 
  k \rightarrow k - \e ,
\end{eqnarray}
the momentum $p_\mu$ in Eq.~(\ref{eq:Bp}) is unchanged.
Reparameterization invariance requires the effective Lagrangian to be
invariant under such transformations and thus ensures the
theory is Lorentz invariant to a given order in $1/M_B$.  Furthermore,
utilizing RI has 
non-trivial consequences as it connects operators of different orders
in the $1/M_B$ expansion and thereby exactly fixes the coefficients of some of the
higher dimensional operators with respect to the lower ones.  We find
that the fixed coefficient Lagrangian is 
\begin{equation}
  \mathfrak{L} = 
    - \left( \Bbar \frac{D_\perp^2}{2 M_B} B \right)
    + \left( \Tbar {}^\mu \frac{D_\perp^2}{2 M_B} T_\mu \right)
   + \cH \left[ \left( \Tbar {}^\mu \frac{i \loarrow D \cdot S}{M_B}
       \vit \cdot A \, T_\mu \right) - \left( \Tbar {}^\mu \, \vit\cdot A \frac{S\cdot i \roarrow D}
       {M_B} T_\mu \right) \right]
\label{eq:fixed}
\end{equation}
where $D_\perp^2 = D^2 - (\vit \cdot D)^2$ and we have kept only terms relevant to 
the calculation of decuplet masses.

The Lagrangian contains additional $\cO (q^2)$ operators as well
as $\cO (q^4)$ operators which are invariant under the $SU(3)$ chiral
transformations.  In part, these operators absorb some effects 
of the unphysical, off-shell degrees of freedom \cite{Hemmert:1997ye}.
The operators relevant to the self energy of the
decuplet baryons are
\begin{eqnarray}
\mathfrak{L} &=& 
\frac{1}{4 \pi f} 
\Bigg[
t_1^A \, \ol T {}^{kji}_\mu (A_\nu A^\nu)_{i}{}^{i'} T^\mu_{i'jk} 
+ 
t_2^A \, \ol T {}^{kji}_\mu (A_\nu)_{i}{}^{i'} (A^\nu)_{j}{}^{j'} T^\mu_{i'j'k} 
+ 
t_3^A \, \left( \ol T_\mu T^\mu \right) \tr ( A_\nu A^\nu )
\notag \\
&& + 
t_1^{\tilde{A}} \, \ol T {}^{kji}_\mu (A^\mu A_\nu)_{i}{}^{i'} T^\nu_{i'jk} 
+ 
t_2^{\tilde{A}} \, \ol T {}^{kji}_\mu (A^\mu)_{i}{}^{i'} (A_\nu)_{j}{}^{j'} T^\nu_{i'j'k} 
+ 
t_3^{\tilde{A}} \, \left( \ol T_\mu T^\nu \right) \tr ( A^\mu A_\nu )
\notag \\
&& + 
t_1^{vA} \, \ol T {}^{kji}_\mu (v \cdot A \, v \cdot A)_{i}{}^{i'} T^\mu_{i'jk} 
+ 
t_2^{vA} \, \ol T {}^{kji}_\mu (v \cdot A)_{i}{}^{i'} (v \cdot A)_{j}{}^{j'} T^\mu_{i'j'k} 
+ 
t_3^{vA} \, \left( \ol T_\mu T^\mu \right) \tr ( v \cdot A \, v \cdot A )
\notag \\
&& +
t_1^M \, \ol T {}^{kji}_\mu (\cM \cM)_{i}{}^{i'} T^\mu_{i'jk} 
+ 
t_2^M  \, \ol T {}^{kji}_\mu (\cM)_{i}{}^{i'} (\cM)_{j}{}^{j'} T^\mu_{i'j'k} 
+
t_3^M  \, \left( \ol T_\mu T^\mu \right) \tr (\cM \cM)
\notag \\
&& +
t_4^M \, \left( \ol T_\mu \cM T^\mu \right) \tr (\cM)
+
t_5^M  \, \left( \ol T_\mu T^\mu \right) \tr(\cM) \, \tr(\cM)
\Bigg]
\label{eq:hdo}.
\end{eqnarray}
All the LECs $t^{A}_i$, $t^{\tilde{A}}_i$, $t^{vA}_i$, and $t^M_i$ are dimensionless.  
In principle, additional $1/M_B$ operators with the same chiral
symmetry properties as those contained in Eq.~(\ref{eq:hdo}) can be
generated.  However, these $1/M_B$ operators do not have their
coefficients constrained by RI, therefore they shall be absorbed into
the definition of the various $t^{A}_i$, $t^{\tilde{A}}_i$, $t^{vA}_i$, and $t^M_i$.  
Since the flavor, spin and Lorentz structure of these omitted $1/M_B$ operators is identical 
to those above, we are guaranteed that the values absorbed in the LECs of Eq.~\eqref{eq:hdo} 
remain the same for all processes and thus the determination of the above LECs is in that sense universal.

Including the decuplet fields in \CPT\ requires additional operators involving $\D / \L_\chi$
because $\D$ is a chiral singlet.  Thus arbitrary functions $f(\D / \L_\chi)$ can
multiply any term in the Lagrangian without changing the
properties under chiral transformations.  To the order we are working, 
all constants in the calculation must be arbitrary polynomial functions of
$\D/\L_\chi$, and expanded out to the appropriate order.  For example
\begin{equation}
  \gm \rightarrow \gm\left(\frac{\D}{\L_\chi}\right)
      = \gm\left[1+\gamma_1 \, \frac{\D}{\L_\chi} 
        + \gamma_2 \, \frac{\D^2}{\L_\chi^2} 
        +\cO\left(\frac{\D^3}{\L_\chi^3}\right)
        \right]\, .
\end{equation}
Furthermore at this order the decuplet-octet mass splitting in the chiral limit is a polynomial 
function of $\D/\Lambda_\chi$. 
We shall not explicitly write down the operators that contribute to the $\Delta$ dependence  
because determination of their LECs requires the ability to vary $\D$. 
Additionally, the LECs are also implicit functions of $\mu$, to absorb
the divergences arising from the loop integrals.

\section{Decuplet masses in \CPT} \label{CPT}
The masses of the octet and decuplet baryons have been investigated considerably in \CPT\ 
\cite{Jenkins:1992ts,Lebed:1994yu,Lebed:1994gt,Banerjee:1995bk,Borasoy:1997bx,Frink:2004ic,Walker-Loud:2004hf}.
Here we calculate the masses of the decuplet baryons to NNLO in \CPT. 
The mass of the $i^{th}$ decuplet baryon in the chiral expansion can be written as
\begin{eqnarray}
     M_{T_i} = M_0 \left(\mu \right) +  M_{T_i}^{(1)}\left(\mu \right)
                + M_{T_i}^{(3/2)}\left(\mu \right)
                + M_{T_i}^{(2)}\left(\mu \right) + \ldots
\label{eq:massexp}
\end{eqnarray}
Here, $M_0 \left(\mu \right)$ is the renormalized mass of the decuplet
baryons in the chiral limit which is independent of $m_Q$ and also of the $T_i$.
$M_{T_i}^{(n)}$ is the contribution to the 
$i^{th}$ decuplet baryon of the order $m_Q^{(n)}$, and $\mu$ is the
renormalization scale.  For this calculation we use dimensional 
regularization with a modified minimal subtraction ($\ol{\rm   MS}$) 
scheme.

\begin{figure}[tb]
  \centering
  \includegraphics[width=0.4\textwidth]{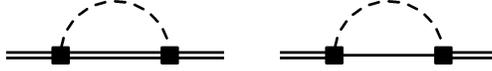}%
  \caption{
    One loop graphs which give contributions to $M_{T_i}^{(3/2)}$.  The
    single, double and dashed lines correspond to octet baryons,
    decuplet baryons and mesons, respectively.  The filled squares
    denote the axial coupling given in Eq.~(\ref{eq:leadlag}).
  }
  \label{fig:NLO}
\end{figure}

\begin{figure}[tb]
  \centering
  \includegraphics[width=0.6\textwidth]{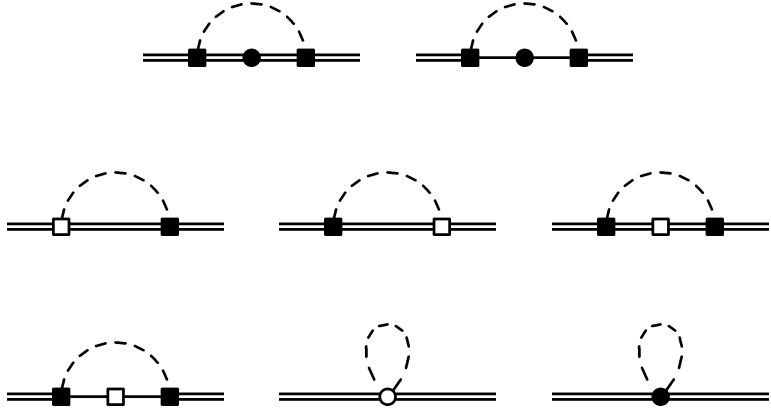}%
  \caption{
    One-loop graphs which give contributions to $M_{T_i}^{(2)}$.   
    Single, double and dashed lines correspond to octet baryons,
    decuplet baryons and mesons, respectively.  The filled squares and
    filled circles denote axial couplings and insertions of $\cM$ given
    in Eq.~(\ref{eq:leadlag}).  The empty squares denote insertions of
    fixed $1/M_B$ operators from Eq.~(\ref{eq:fixed}), and the empty
    circles denote two-baryon-axial couplings defined in
    Eq.~(\ref{eq:hdo}).
  }
  \label{fig:NNLO}
\end{figure}

In calculating the masses, the leading dependence upon $m_Q$ arises
from the terms the Lagrangian Eq. (\ref{eq:leadlag}) 
with coefficients $\gm$ and $\smt$. 
The $\cO (m_Q^{3/2})$ contributions arise from the one-loop diagrams shown
in Fig.~\ref{fig:NLO},
which are formed from the operators in the Lagrangian with coefficients
$\cH$ and $\cC$.  The $\cO (m_Q^2)$ contributions arise from the
one-loop diagrams shown in Fig.~\ref{fig:NNLO},
from the tree-level contributions of the operators with coefficients,
$t^M_i$, and from the NLO wave-function corrections.

\begin{table}
\caption{The tree level coefficients $m_T$, $(m^2)_T$, and $(mm')_T$ in \CPT\ and \PQCPT\ for decuplet states $T$.}
\begin{tabular}{l | c c c }
 & $m_T$ & $ \phantom{spac} (m^2)_T \phantom{spac}$ & $(mm')_T$ \\
\hline
$\D$       
             &  $3 m_u$ & $3 m_u^2$  & $3m_u^2$  \\
$\Sigma^*$ 
 	     &  $2 m_u + m_s$ & $2 m_u^2 + m_s^2$  & $m_u^2 + 2 m_u m_s$  \\
$\Xi^*$    
      	     &  $m_u + 2 m_s$ & $m_u^2 + 2 m_s^2$  & $2 m_u m_s + m_s^2$  \\
$\Omega^-$ 
             &  $3 m_s$ & $3 m_s^2$  & $3 m_s^2$  \\
\end{tabular}
\label{t:mT}
\end{table}

We find that the leading-order contributions to the decuplet masses are
\begin{equation}
M^{(1)}_T = \frac{2}{3} \, \gamma_M \, m_T  - 2 \ol \sigma_M \, \tr (m_Q),
\label{eq:MLO}
\end{equation}
where the coefficients $m_T$ appear in Table \ref{t:mT}.
The next-to-leading order contributions are
\begin{equation}
M^{(3/2)}_T = -\frac{5 \cH^2}{72 \pi f^2} \sum_\phi A_\phi^T \, m_\phi^3 
- \frac{\cC^2}{(4 \pi f)^2} \sum_\phi B_\phi^T  \, \cF (m_\phi,-\D,\mu)
,
\label{eq:MNLO}
\end{equation}
where the function $\cF$ is defined by
\begin{eqnarray}
\cF (m,\d,\mu) &=& (m^2 - \d^2) 
\left[ \sqrt{\d^2 - m^2}
\log 
\left( 
\frac{\d - \sqrt{\d^2 - m^2 + i \varepsilon}}
{\d + \sqrt{\d^2 - m^2 + i \varepsilon}} \right)
- 
\d \log \left( \frac{m^2}{\mu^2} \right)
\right] \notag \\
&& \phantom{sp} - \frac{1}{2} \d m^2 \log \left( \frac{m^2}{\mu^2} \right)
\label{eqn:F}
.\end{eqnarray}
In Eq.~\eqref{eq:MNLO} the sum on $\phi$ runs over loop mesons with the mass $m_\phi$.
The coefficients $A_\phi^T$ and $B_\phi^T$ above are the sums of squares of Clebsch-Gordon coefficients.
The numerical values of these coefficients depend on the decuplet state $T$ and are listed in 
Table \ref{t:QCD-AB} for loop mesons with mass $m_\phi$.

\begin{table}
\caption{The coefficients $A^T_\phi$ and $B^T_\phi$ in \CPT\ for decuplet states $T$.}
\begin{tabular}{l | c c c | c c c }
 & \multicolumn{3}{c|}{$A^T_\phi \phantom{ap}$} & \multicolumn{3}{c}{$B^T_\phi$ \phantom{sp}} \\
$\phi$ & $\quad \pi \quad$ & $\quad K \quad $ 
    & $\quad \eta \quad $ & $ \quad \pi \quad$ & $ \quad K \quad$ 
    & $\quad \eta \quad$ \\
\hline
$\D$       &  $\frac{5}{6}$ & $\frac{1}{3}$  & $\frac{1}{6}$  
           &  $\frac{2}{3}$ & $\frac{2}{3}$  & $0$  \\

$\Sigma^*$ &  $\frac{4}{9}$ & $\frac{8}{9}$  & $0$  
           &  $\frac{5}{9}$ & $\frac{4}{9}$  & $\frac{1}{3}$\\

$\Xi^*$    &  $\frac{1}{6}$ & $1$  & $\frac{1}{6}$  
           &  $\frac{1}{3}$ & $\frac{2}{3}$  & $\frac{1}{3}$\\

$\Omega^-$ &  $0$ & $\frac{2}{3}$  & $\frac{2}{3}$  
           &  $0$ & $\frac{4}{3}$  & $0$\\
 
\end{tabular}
\label{t:QCD-AB}
\end{table}

Finally the next-to-next-to-leading contributions to the decuplet mass are

\begin{table}
\caption{The coefficients $C_\phi^T$ and $\ol C_\phi$ in \CPT. Coefficients are
listed for the decuplet states $T$ and grouped into contributions from loop mesons
with mass $m_\phi$. $\ol C_\phi$ is identical for each decuplet state.}
\begin{tabular}{l | c c c }
$C_\phi^T$  & $\quad \quad \pi \quad\quad $ & $\quad \quad K \quad\quad $ & $\quad\quad \eta \quad\quad $ \\
\hline
$\D$       
           &  $3 m_u$ & $m_u + m_s$  & $\frac{1}{3} m_u$  \\

$\Sigma^*$ 
           &  $2 m_u$ & $\frac{4}{3}(m_u + m_s)$  & $\frac{2}{9}(m_u + 2 m_s)$\\

$\Xi^*$    
           &  $m_u$ & $\frac{5}{3} (m_u + m_s)$  & $\frac{1}{9}( m_u + 8 m_s)$\\

$\Omega^-$ 
           &  $0$ & $2 (m_u + m_s)$  & $\frac{4}{3} m_s$\\
\hline
\hline
$\ol C_\phi$ 
	   & $6 m_u$ & $4(m_u + m_s)$ & $\frac{2}{3} (m_u + 2 m_s)$ 
 \end{tabular}
\label{t:QCD-C}
\end{table}

\begin{table}
\caption{The coefficients $D_\phi^T$ and $\ol D_\phi$ in \CPT. Coefficients are
listed for the decuplet states $T$ and grouped into contributions from loop mesons
with mass $m_\phi$. $\ol D_\phi$ is identical for each decuplet state.}
\begin{tabular}{l | c c c }
$D_\phi^T$  & $\quad \quad \pi \quad\quad $ & $\quad \quad K \quad\quad $ & $\quad\quad \eta \quad\quad $ \\
\hline
$\D$       
           &  $\frac{3}{2}$ & $1$  & $\frac{1}{6}$  \\

$\Sigma^*$ 
           &  $1$ & $\frac{4}{3}$  & $\frac{1}{3}$\\

$\Xi^*$    
           &  $\frac{1}{2}$ & $\frac{5}{3}$  & $\frac{1}{2}$\\

$\Omega^-$ 
           &  $0$ & $2$  & $\frac{2}{3}$\\
\hline
\hline
$\ol D_\phi$ 
	   & $3$ & $4$ & $1$ 
 \end{tabular}
\label{t:QCD-D}
\end{table}

\begin{eqnarray}
M_T^{(2)} &=& 
(Z - 1) M_T^{(1)} 
\notag \\ 
&& + 
\frac{1}{4 \pi f} 
\left\{
\frac{1}{3} t_1^M \, (m^2)_T 
+ 
\frac{1}{3} t_2^M \, (m m')_T 
+ 
t_3^M \, \tr(m_Q^2) 
+ 
\frac{1}{3} t_4^M \,  m_T  \, \tr (m_Q)
+
t_5^M \, [\tr (m_Q)]^2
\right\}
\notag \\
&& -
\frac{2 \,\gamma_M}{(4 \pi f)^2} \sum_\phi C_\phi^T \, \cL(m_\phi,\mu) 
+ 
\frac{2 \, \ol \sigma_M}{(4 \pi f)^2} \sum_\phi \ol C_\phi \, \cL(m_\phi,\mu)
\notag \\
&& + 
\frac{1}{(4 \pi f )^3} 
\sum_\phi 
\left( 
t^A_1 \,  D_\phi^T + t^A_2 \, E_\phi^T + t^A_3 \, \ol D_\phi 
\right)
\ol \cL (m_\phi, \mu)
\notag \\
&& + 
\frac{1}{4 (4 \pi f )^3} 
\sum_\phi 
\left[ 
(t^{\tilde{A}}_1 + t_1^{vA}) D_\phi^T 
+ 
(t^{\tilde{A}}_2 + t_2^{vA}) E_\phi^T 
+ 
(t^{\tilde{A}}_3 + t_3^{vA}) \ol D_\phi 
\right]
\left[
\ol \cL (m_\phi, \mu)
-
\frac{1}{2} m_\phi^4
\right]
\notag \\
&& -
\frac{5}{8} \frac{\cH^2}{(4\pi f)^2 M_B} 
\sum_\phi A_\phi^T 
\left[ 
\ol \cL(m_\phi,\mu)
+ \frac{19}{10} m_\phi^4
\right]
- 
\frac{15}{16} \frac{\cC^2}{(4 \pi f)^2 M_B}  
\sum_\phi B_\phi^T  
\left[ 
\ol \cL(m_\phi,\mu)
- \frac{1}{10} m_\phi^4
\right]
\notag \\
&& - 
\frac{10 \cH^2 \ol \sigma_M \tr(m_Q) }{3 (4 \pi f)^2} 
\sum_\phi A_\phi^T 
\left[ 
\cL(m_\phi,\mu) + \frac{26}{15} m_\phi^2
\right]
- 
\frac{3 \cC^2 \sigma_M \tr(m_Q)  }{ (4 \pi f)^2} 
\sum_\phi B_\phi^T  \, \cJ (m_\phi,-\D,\mu)
\notag \\
&& + 
\frac{10 \cH^2 \gamma_M}{9 (4 \pi f)^2} 
\sum_\phi F_\phi^T 
\left[ \cL(m_\phi,\mu) + \frac{26}{15} m_\phi^2
\right] 
- 
\frac{3 \cC^2 }{2 (4 \pi f)^2} 
\sum_\phi G_\phi^T  \, \cJ (m_\phi,-\D,\mu)
\label{eq:MNNLO}
.\end{eqnarray}
Here the tree-level contributions are expressed in terms of coefficients $m_T$,  $(m^2)_T$, and $(m m')_T$ 
which are listed in Table \ref{t:mT}. 
Above the wavefunction renormalization $Z$ is given by
\begin{equation}
Z - 1 
= 
- \frac{5 \cH^2}{3 (4 \pi f)^2} \sum_\phi A_\phi^T 
\left[ \cL(m_\phi,\mu) + \frac{26}{15} m_\phi^2 \right] 
- \frac{3 \cC^2}{2 (4 \pi f)^2} \sum_\phi B_\phi^T  \, \cJ (m_\phi,-\D,\mu)
\label{eq:Z}
.\end{equation}
The above equations \eqref{eq:MNNLO} and \eqref{eq:Z} also employ abbreviations for non-analytic functions arising from loop contributions. These
functions are
\begin{align}
\cL (m,\mu) &= m^2 \log \frac{m^2}{\mu^2} \\
\ol \cL (m,\mu) &= m^4 \log \frac{m^2}{\mu^2} \\
\cJ (m,\d,\mu) &= (m^2 - 2 \d^2) \log \frac{m^2}{\mu^2} + 2 \d \sqrt{\d^2 - m^2} \log
\left(
\frac{\d - \sqrt{\d^2 - m^2 + i \varepsilon}}{\d + \sqrt{\d^2 - m^2 + i \varepsilon}}
\right) 
.\end{align}
As in Eq.~\eqref{eq:MNLO}, the sums over $\phi$ in Eqs.~\eqref{eq:MNNLO} and \eqref{eq:Z} 
run through loop mesons with mass $m_\phi$. The coefficients in these sums are themselves the sums of squares of Clebsch-Gordon
coefficients and/or quark mass insertions. These are listed in Tables \ref{t:QCD-C}--\ref{t:QCD-G}.
To be clear, the above expressions Eqs.~\eqref{eq:MLO}, \eqref{eq:MNLO}, and \eqref{eq:MNNLO} are quark mass
expressions, i.e.~, the meson masses in these expressions are merely replacements for the quark masses
via Eq.~\eqref{eq:mesonmass}. To utilize these expressions for chiral extrapolations of lattice QCD data, one must 
write the quark masses in terms of physical meson masses (as determined on the lattice).
To work consistently to $\cO(m_q^2)$, one needs to use the NLO expressions for the meson masses in the LO results
in Eq.~\eqref{eq:MLO}

\begin{table}
\caption{The coefficients $E_\phi^T$ in \CPT. Coefficients are
listed for the decuplet states $T$ and grouped into contributions from loop mesons
with mass $m_\phi$.}
\begin{tabular}{l | c c c }
$E_\phi^T$  & $\quad \quad \pi \quad\quad $ & $\quad \quad K \quad\quad $ & $\quad\quad \eta \quad\quad $ \\
\hline
$\D$       
           &  $\frac{1}{2}$ & $0$  & $\frac{1}{6}$  \\

$\Sigma^*$ 
           &  $\frac{1}{6}$ & $\frac{2}{3}$  & $-\frac{1}{6}$\\

$\Xi^*$    
           &  $0$ & $\frac{2}{3}$  & $0$\\

$\Omega^-$ 
           &  $0$ & $0$  & $\frac{2}{3}$\\
\end{tabular}
\label{t:QCD-E}
\end{table}

\begin{table}
\caption{The coefficients $F_\phi^T$ in \CPT. Coefficients are
listed for the decuplet states $T$ and grouped into contributions from loop mesons
with mass $m_\phi$.}
\begin{tabular}{l | c c c }
$F_\phi^T$  & $\quad \quad \pi \quad\quad $ & $\quad \quad K \quad\quad $ & $\quad\quad \eta \quad\quad $ \\
\hline
$\D$       
           &  $\frac{5}{2} m_u$ & $\frac{1}{3} (2 m_u + m_s)$  & $\frac{1}{2} m_u$  \\

$\Sigma^*$ 
           &  $\frac{4}{9} (2 m_u + m_s)$ & $\frac{8}{9} (2 m_u + m_s)$  & $0$\\

$\Xi^*$    
           &  $\frac{1}{6}(m_u + 2 m_s)$ & $\frac{1}{3}(4 m_u + 5 m_s)$  & $\frac{1}{6}(m_u + 2 m_s)$\\

$\Omega^-$ 
           &  $0$ & $\frac{2}{3} ( m_u + 2 m_s)$  & $2 m_s$\\
\end{tabular}
\label{t:QCD-F}
\end{table}

\begin{table}
\caption{The coefficients $G_\phi^T$ in \CPT. Coefficients are
listed for the decuplet states $T$ and grouped into contributions from loop mesons
with mass $m_\phi$.}
\begin{tabular}{l | c c c }
$G_\phi^T$  & $\quad \quad \pi \quad\quad $ & $\quad \quad K \quad\quad $ & $\quad\quad \eta \quad\quad $ \\
\hline
$\D$       
&  $\frac{4}{3}m_u (\a_M + \b_M)$ & $\frac{2}{9} [m_u ( 5 \a_M + 2 \b_M)$  & $0$  \\
&                                 & $\qquad + m_s(\a_M + 4 \b_M)]$                               &      \\

\hline

$\Sigma^*$ 
&  $\frac{1}{27}[m_u( 19 \a_M + 22 \b_M)$ 
& $\frac{2}{27}[m_u(7 \a_M + 10 \b_M)$  
&  $\frac{1}{9}[m_u(5 \a_M + 2 \b_M)$\\

& $\qquad + m_s (11 \a_M + 8 \b_M)]$
& $\qquad + m_s (5 \a_M + 2 \b_M)]$
& $\qquad + m_s (\a_M + 4 \b_M)]$ \\

\hline

$\Xi^*$    
&  $\frac{1}{9}[m_u(\a_M + 4 \b_M)$ 
&  $\frac{4}{9}[2 m_u (\a_M + \b_M)$ 
&  $\frac{1}{9}[m_u(\a_M + 4 \b_M)$\\

& $\qquad + m_s (5 \a_M + 2 \b_M)]$
& $\qquad + m_s (\a_M + \b_M)]$
& $\qquad + m_s (5 \a_M + 2 \b_M)]$ \\

\hline

$\Omega^-$ 
&  $0$ 
& $\frac{4}{9}[m_u (\a_M + 4 \b_M)$  
& $0$\\

& 
& $\qquad + m_s (5 \a_M + 2 \b_M)]$ 
& \\

\end{tabular}
\label{t:QCD-G}
\end{table}

\section{\PQCPT} \label{PQintro}

In PQQCD, the quark part of the Lagrangian is
\begin{equation}
\mathfrak{L} = \sum_{j,k=1}^9 \ol{Q}{}^{\hskip 0.2em j} \left(
  i\Dslash - m_Q \right)_j^{\hskip 0.3em k} Q_k
.\label{eq:pqqcdlag}
\end{equation}
This differs from the $SU(3)$ Lagrangian of QCD by the
inclusion of six extra quarks; three bosonic ghost quarks, ($\tilde u,
\tilde d, \tilde s$), and three fermionic sea quarks, ($j, l, r$), in addition
to the light physical quarks ($u, d, s$).  The
nine quark fields transform in the fundamental representation of the
graded $SU(6|3)$ group.  They have been accommodated in the nine-component vector
\begin{equation}
  Q^\dagger = (u, d, s, j, l, r, \tilde{u}, \tilde{d}, \tilde{s})
.\end{equation}
The quark fields obey the graded equal-time commutation relation
\begin{equation}
Q^\a_i(\mathbf x) Q^{\beta \dagger}_j(\mathbf y) -
(-1)^{\eta_i \eta_j} Q^{\b \dagger}_j(\mathbf y) Q^\a_i(\mathbf x) =
\d^{\a \b} \d_{ij} \d^3 (\mathbf {x-y}),
\end{equation}
where $\a, \beta$ are spin and $i,j$ are flavor indices.
Analogous graded equal-time commutation relations can be written for
two $Q$'s and two $Q^\dagger$'s.  The grading factors
\begin{equation}
   \eta_k
   = \left\{ 
       \begin{array}{cl}
         1 & \text{for } k=1,2,3,4,5,6 \\
         0 & \text{for } k=7,8,9
       \end{array} 
     \right.
\end{equation}
take into account the different fermionic and bosonic statistics of
the quark fields.  In the isospin limit the quark mass matrix of
$SU(6|3)$ is given by
\begin{equation}
  m_Q = \diag(m_u, m_u, m_s, m_j, m_j, m_r, m_u, m_u, m_s).
\end{equation}
Because the ghost quark masses are identically equal to the valence quark masses
there is an exact cancellation in the path integral between the
valence quark determinant and the ghost quark determinant.   The sea
quark determinant is unaffected. Thus in PQQCD, one has the ability to vary the
valence and sea quark masses independently.  
QCD is recovered in the limit $m_j \rightarrow m_u$ and $m_r \rightarrow m_s$.

\subsection{Pseudo-Goldstone Mesons}
For massless quarks, the theory corresponding to the Lagrangian in
Eq. (\ref{eq:pqqcdlag}) has a 
graded $SU(6|3)_L \otimes SU(6|3)_R \otimes U(1)_V$ symmetry which is
assumed to be spontaneously broken down to $SU(6|3)_V \otimes U(1)_V$
in analogy with QCD.  The effective low-energy theory obtained by
perturbing about the physical vacuum state of PQQCD is PQ$\chi$PT.
The result is 80~pseudo-Goldstone mesons with dynamics described 
at leading order in the chiral expansion by the Lagrangian
\begin{equation}
  \mathfrak{L} =
    \frac{f^2}{8} \str \left(
      \partial^\mu \S^\dagger \partial_\mu \S \right)
      + \l  \, \str \left( m_q \S^\dagger + m_q^\dagger \S \right)
           +\a_\Phi \partial^\mu \Phi_0 \partial_\mu \Phi_0
           - m_0^2 \Phi_0^2,
\label{eq:pqbosons}
\end{equation}
where
\begin{equation}
  \S = \exp \left( \frac{2 i \Phi}{f} \right) = \xi^2
,\end{equation}
and the meson fields appear in
\begin{equation}
    \Phi =
    \begin{pmatrix}
      M & \chi^\dagger\\
      \chi & \tilde M\\
    \end{pmatrix}.
\end{equation}
The operation $\str( )$ in Eq. (\ref{eq:pqbosons}) is the
supertrace over flavor indices.  The quantities $\a_\Phi$ and
$m_0$ are non-vanishing in the chiral limit.  $M$ and $\tilde M$ are
matrices containing bosonic mesons (with quantum numbers of $q \bar{q}$ pairs and 
$\tilde{q} \bar{\tilde{q}}$ pairs, respectively), while $\chi$ and $\chi^\dagger$
are matrices containing fermionic mesons (with quantum numbers of $\tilde q \bar{q}$
pairs and $q \bar{\tilde{q}}$ pairs, respectively).
The upper $3 \times 3$ block of $M$ is the usual octet of
pseudo-scalar mesons and the remaining components are mesons formed
with one or two sea quarks, see e.g.~\cite{Chen:2001yi}.

  The flavor singlet field is defined to be $\Phi_0 = {\rm str}( \Phi ) /
{\sqrt 6}$.  PQQCD has a strong axial anomaly $U(1)_A$
and therefore the mass of the singlet field $m_0$ can be taken to be
the order of the chiral symmetry breaking scale, $m_0 \rightarrow
\Lambda_\chi$ \cite{Sharpe:2000bn}.  In this limit, the $\eta$
two-point correlation functions deviate from their form in \CPT.
For $a,b = u,d,s,j,l,r,\tilde u, \tilde d, \tilde s$, 
the $\eta_a \eta_b$ propagator with $2+1$ sea-quarks at leading
order is
\begin{equation}
{\cal G}_{\eta_a \eta_b} =
        \frac{i \epsilon_a \delta_{ab}}{q^2 - m^2_{\eta_a} +i\epsilon}
        - \frac{i}{3} \frac{\epsilon_a \epsilon_b \left(q^2 - m^2_{jj}
            \right) \left( q^2 - m^2_{rr} \right)}
            {\left(q^2 - m^2_{\eta_a} +i\epsilon \right)
             \left(q^2 - m^2_{\eta_b} +i\epsilon \right)
             \left(q^2 - m^2_X +i\epsilon \right)}\, ,
\end{equation}
where
\begin{equation}
\epsilon_a = (-1)^{1+\eta_a}
.\end{equation}
The mass $m_{xy}$ is the mass of a meson composed of (anti)-quarks
of flavor $x$ and $y$, while the mass $m_X$ is defined as $m_X^2 =
\frac{1}{3}\left(m^2_{jj} + 2   m^2_{rr}\right)$.  The singlet propagator 
can be conveniently rewritten as
\begin{equation}
{\cal G}_{\eta_a \eta_b} =
         \e_a \d_{ab} P_a +
         \e_a \e_b {\cal H}_{ab}\left(P_a,P_b,P_X\right),
\end{equation}
where
\begin{eqnarray}
     P_a &=& \frac{i}{q^2 - m^2_{\eta_a} +i\e}\ ,\ 
     P_b = \frac{i}{q^2 - m^2_{\eta_b} +i\e}\ ,\ 
     P_X = \frac{i}{q^2 - m^2_X +i\e} \,
\nonumber\\
\nonumber\\
\nonumber\\
     {\cal H}_{ab}\left(A,B,C\right) &=& 
           -\frac{1}{3}\left[
             \frac{\left( m^2_{jj}-m^2_{\eta_a}\right)
                   \left( m^2_{rr}-m^2_{\eta_a}\right)}
                  {\left( m^2_{\eta_a} - m^2_{\eta_b}\right)
                   \left( m^2_{\eta_a} - m^2_X\right)}
                 A
            -\frac{\left( m^2_{jj}-m^2_{\eta_b}\right)
                   \left( m^2_{rr}-m^2_{\eta_b}\right)}
                  {\left( m^2_{\eta_a} - m^2_{\eta_b}\right)
                   \left( m^2_{\eta_b} - m^2_X\right)}
                 B \right.\, 
\nonumber\\
&&\qquad\quad\left.
            +\frac{\left( m^2_X-m^2_{jj}\right)
                   \left( m^2_X-m^2_{rr}\right)}
                  {\left( m^2_X-m^2_{\eta_a}\right)
                   \left( m^2_X-m^2_{\eta_b}\right)}
                 C\ \right].
\label{eq:Hfunction}
\end{eqnarray}

\subsection{Baryons}

In PQ$\chi$PT the baryons are composed of three quarks $Q_i Q_j Q_k$
where $i-k$ can be valence, sea or ghost quarks. 
One decomposes the irreducible representations of $SU(6|3)_V$ into
irreducible representations of $SU(3)_{val} \otimes SU(3)_{sea}
\otimes SU(3)_{ghost} \otimes U(1)$.  The method for including the
octet and decuplet baryons into PQ$\chi$PT is to use the interpolating
field~\cite{Labrenz:1996jy,Chen:2001yi}
\begin{equation}
  \cB_{ijk}^\g \sim
    \left(Q_i^{\a,a} Q_j^{\beta,b} Q_k^{\g,c}-Q_i^{\a,a}
    Q_j^{\g,c} Q_k^{\beta,b}\right)
    \epsilon_{abc}(C\g_5)_{\a \beta}.
\end{equation} 
Under the interchange of flavor indices, one
finds~\cite{Labrenz:1996jy}
\begin{equation}
\cB_{ijk} = (-)^{1+\eta_j \eta_k}\cB_{ikj}
           \quad {\rm and} \quad
  \cB_{ijk} + (-)^{1+\eta_i \eta_j}\cB_{jik}
                 + (-)^{1+\eta_i \eta_j + \eta_j \eta_k + \eta_i
                   \eta_k} \cB_{kji}
                 =0.
\end{equation}
We require that $\cB_{ijk} = B_{ijk}$, defined in
Eq.~(\ref{eq:3indexB}), when the indices, $i,j,k$ are restricted to
$1-3$.  Thus the octet baryons are contained as an $(\mathbf{8,1,1})$ 
of $SU(3)_{val} \otimes SU(3)_{sea} \otimes SU(3)_{ghost} \otimes
U(1)$ in the $\mathbf{240}$ representation.  In addition to the
conventional octet baryons composed of valence quarks, $\cB_{ijk}$
also contains baryon fields composed of sea and ghost quarks.  In this
work we only need states of the $\mathbf{240}$  which contain at most one
sea or ghost quark, and these states have been explicitly constructed
in~\cite{Chen:2001yi}.

Similarly, one can construct the spin-$\frac{3}{2}$ baryons
which make up the $\mathbf{138}$, and have an interpolating
field
\begin{equation}
  \cT_{ijk}^{\a,\mu} \sim
      \left( Q_i^{\a,a} Q_j^{\beta,b} Q_k^{\g,c}
            +Q_i^{\beta,b} Q_j^{\g,c} Q_k^{\a,a}
            +Q_i^{\g,c} Q_j^{\a ,a} Q_k^{\beta,b}
      \right)
         \epsilon_{abc} \left(C\g^\mu \right)_{\beta \g}.
\end{equation}
Under the interchange of flavor indices, one finds that
\begin{equation}
\cT_{ijk} = (-)^{1 + \eta_i \eta_j}\cT_{jik} = 
                 (-)^{1 + \eta_j \eta_k}\cT_{ikj}\, .
\end{equation}
We require that $\cT_{ijk} = T_{ijk}$, when the indices $i,j,k$ are
restricted to $1-3$.  Under $SU(3)_{val} \otimes SU(3)_{sea} \otimes
SU(3)_{ghost} \otimes U(1)$ they transform as a $(\mathbf{ 10,1,1})$.
In addition to the conventional decuplet resonances composed of
valence quarks, $\cT_{ijk}$ contains fields with sea and ghost
quarks. As with the $\mathbf{240}$, for our calculation the required 
states of the $\mathbf{138}$ have been constructed in~\cite{Chen:2001yi}.

To write down the PQ$\chi$PT Lagrangian, we must also include the
appropriate grading factors in the contraction of flavor indices. These 
are included in the $()$ notation as was originally defined 
in~\cite{Labrenz:1996jy,Chen:2001yi}
The leading order \PQCPT\ Lagrangian is given by
\begin{eqnarray}
  \mathfrak{L} &=&
    \left( \cBbar\ i\vD\ \cB \right)\ 
    +\ 2\am \left( \cBbar \cB \cM \right)\ 
    +\ 2\bm \left( \cBbar \cM \cB \right)\ 
    +\ 2\smb \left( \cBbar \cB \right) \str(\cM)
\nonumber\\
 && -
    \left(\cTbar {}^{\mu}\left[\ i\vD - \Delta\ \right] \cT_\mu \right)\ 
    +\ 2 \gm \left(\cTbar{}^\mu \cM \cT_\mu \right)\ 
    -\ 2 \smt \left(\cTbar{}^\mu \cT_\mu \right) \str (\cM)
\nonumber\\
 && +
    2\a\left(\cBbar S^\mu \cB A_\mu \right)\ 
    +\ 2\beta \left(\cBbar S^\mu A_\mu \cB \right)\ 
    +\ 2{\cal H} \left( \cTbar {}^\nu S^\mu A_\mu \cT_\nu \right)
\nonumber\\
 && +
    \sqrt{\frac{3}{2}}{\cC} \left[ \left( \cTbar{}^\nu A_\nu \cB
      \right)\ 
    +\ \left( \cBbar A_\nu \cT^\nu \right) \right]
\label{eq:leadlagPQ}\,.
\end{eqnarray} 
The low-energy constants appearing above have the same numerical values as those 
in \CPT.

At higher orders, the situation is quite similar to the \CPT\ case considered
in Section \ref{intro}. 
Recall that at higher orders, the Lagrangian can contain arbitrary functions
of $\D / \L_\chi$. We take this into account by implicitly treating 
the leading order coefficients as functions of $\D / \L_\chi$ expanded out 
to the required order. To enforce Lorentz invariance on the theory, we use RI to generate the 
higher dimensional operators with fixed coefficients. 
In \PQCPT\ the fixed coefficient Lagrangian is given by
\begin{eqnarray}
 \mathfrak{L} &=& 
    - \left( \ol \cB \frac{D_\perp^2}{2 M_B} \cB \right)
    + \left( \ol \cT {}^\mu \frac{D_\perp^2}{2 M_B} \cT_\mu \right)
   + \cH \left[ \left( \ol \cT {}^\mu \frac{i \loarrow D \cdot S}{M_B}
       \vit \cdot A \, \cT_\mu \right) - \left( \ol \cT {}^\mu \, \vit\cdot A \frac{S\cdot i \roarrow D}
       {M_B} \cT_\mu \right) \right].
\notag
\\
&& \label{eq:fixedPQ}
\end{eqnarray}
As in \CPT\ there are additional operators with unfixed coefficients. 
These higher dimensional operators relevant for our calculation are collected in the \PQCPT\ Lagrangian\footnote{%
We have omitted three \PQCPT\ operators of the form $\ol \cT {}^\mu \, [ A_\mu , A_\nu ] \cT^\nu$  and three of the form
$\ol \cT {}^\mu \, [ A_\nu , A_\rho ] S^\nu S^\rho \cT_\mu$ since their contributions
to the decuplet masses vanish to the order we are working. Such operators identically vanish in \CPT.}
\begin{eqnarray}
\mathfrak{L} &=&
\frac{1}{4 \pi f} 
\Bigg[ 
t_1^A \, \ol \cT {}^{kji}_\mu (A_\nu A^\nu)_{i}{}^{i'} \cT^\mu_{i'jk} 
+ 
t_2^A   (-)^{\eta_{i'} (\eta_j + \eta_{j'})}  \ol \cT {}^{kji}_\mu (A_\nu)_{i}{}^{i'} (A^\nu)_{j}{}^{j'} \cT^\mu_{i'j'k} 
+ 
t_3^A \, \left( \ol \cT_\mu \cT^\mu \right) \str ( A_\nu A^\nu )
\notag \\
&& + 
t_1^{\tilde{A}} \, \ol \cT {}^{kji}_\mu (A^\mu A_\nu)_{i}{}^{i'} \cT^\nu_{i'jk} 
+ 
t_2^{\tilde{A}} \, \ol \cT {}^{kji}_\mu (A^\mu)_{i}{}^{i'} (A_\nu)_{j}{}^{j'} \cT^\nu_{i'j'k} 
+ 
t_3^{\tilde{A}} \, \left( \ol \cT_\mu \cT^\nu \right) \tr ( A^\mu A_\nu )
\notag \\
&& + 
t_1^{vA} \, \ol \cT {}^{kji}_\mu (v \cdot A \, v \cdot A)_{i}{}^{i'} \cT^\mu_{i'jk} 
+ 
t_2^{vA} \, \ol \cT {}^{kji}_\mu (v \cdot A)_{i}{}^{i'} (v \cdot A)_{j}{}^{j'} \cT^\mu_{i'j'k} 
+ 
t_3^{vA} \, \left( \ol \cT_\mu \cT^\mu \right) \tr ( v \cdot A \, v \cdot A )
\notag \\
&& +t_1^M \, \ol \cT {}^{kji}_\mu (\cM \cM)_{i}{}^{i'} \cT^\mu_{i'jk} 
+ 
t_2^M  (-)^{\eta_{i'} (\eta_j + \eta_{j'})} \ol \cT {}^{kji}_\mu (\cM)_{i}{}^{i'} (\cM)_{j}{}^{j'} \cT^\mu_{i'j'k} 
\notag \\
&&+
t_3^M  \, \left( \ol \cT_\mu \cT^\mu \right) \tr (\cM \cM)
+
t_4^M \, \left( \ol \cT_\mu \cM \cT^\mu \right) \tr (\cM)
+
t_5^M  \, \left( \ol \cT_\mu \cT^\mu \right) \tr(\cM) \, \tr(\cM)
\Bigg]
.\notag \\
&& 
\label{eq:hdoPQ}
\end{eqnarray}
The coefficients of these operators $t_i^{A}$, $t_i^{\tilde{A}}$, $t_i^{vA}$, and $t_i^{M}$ have the same numerical values as in \CPT.

\section{Decuplet masses in \PQCPT.}\label{PQCPT}
\begin{figure}[t]
  \centering
  \includegraphics[width=0.4\textwidth]{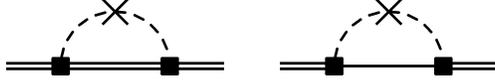}%
  \caption{
    In addition to the one-loop diagrams in Fig.~\ref{fig:NLO},
    $M_{T_i}^{(3/2)}$ also receives contributions from the singlet
    (hairpins) in PQ$\chi$PT.  Single and doubles line correspond
    to $\mathbf {240}$-baryons and $\mathbf {138}$-baryons
    respectively.  The crossed 
    dashed line denotes a hairpin propagator.  The filled squares
    denote the axial coupling given in Eq.~(\ref{eq:leadlagPQ}).
  }
  \label{fig:PQNLO}
\end{figure}

\begin{figure}[b]
  \centering
  \includegraphics[width=0.6\textwidth]{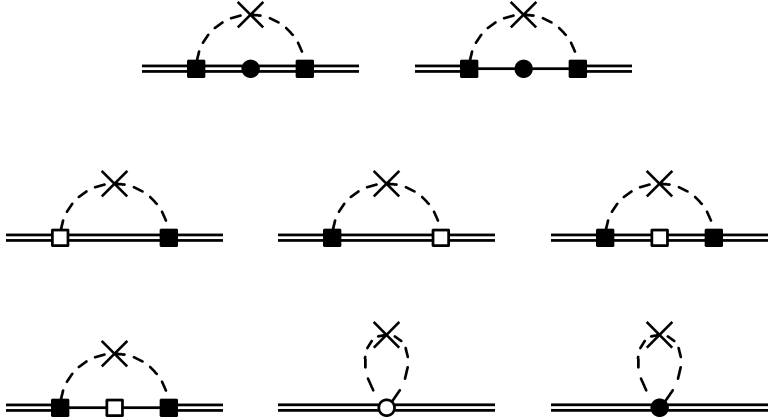}%
  \caption{
    In addition to the one-loop diagrams in Fig.~\ref{fig:NNLO},
    $M_{T_i}^2$ also receives contributions from the singlet
    (hairpins) in PQ$\chi$PT.  Single and double lines correspond
    to $\mathbf {240}$-baryons and $\mathbf {138}$-baryons
    respectively.  The crossed dashed line denotes a hairpin
    insertion.  Filled squares denote the axial coupling given in
    Eq.~(\ref{eq:leadlagPQ}).  Filled circles denote a coupling to
    $\cM$ given in Eq.~(\ref{eq:leadlagPQ}).  Empty squares and
    empty circles denote insertions of fixed $1/M_B$ operators given in
    Eq.~(\ref{eq:fixedPQ}), and insertions of a two-baryon-two-axial
    coupling given in Eq.~(\ref{eq:hdoPQ}) respectively.
  }
  \label{fig:PQNNLO}
\end{figure}

Only the masses of the octet baryons have been investigated in \PQCPT\ 
\cite{Chen:2001yi,Walker-Loud:2004hf}.
Here we calculate the masses of the decuplet baryons to NNLO in \PQCPT. 
The chiral expansion of the decuplet baryon masses in PQ$\chi$PT has
the same form as in $\chi$PT.
\begin{eqnarray}
     M_{T_i} = M_0 \left(\mu \right) +  M_{T_i}^{(1)}\left(\mu \right)
                + M_{T_i}^{(3/2)}\left(\mu \right)
                + M_{T_i}^{(2)}\left(\mu \right) + \ldots
\label{eq:massexp2}
\end{eqnarray}
The contributions at each order are similar to those in Section \ref{CPT}. 
However, we must also include hairpin contributions from the flavor diagonal
propagator, see Fig.~\ref{fig:PQNLO} and Fig.~\ref{fig:PQNNLO}.

\begin{table}
\caption{The coefficients $A^T_\phi$ and $A^T_{\phi\phi'}$ in \PQCPT. Coefficients are
listed for the decuplet states $T$, and for $A^T_\phi$ are grouped into contributions from loop mesons
with mass $m_\phi$, while for $A^T_{\phi\phi'}$ are grouped into contributions from pairs of quark-basis 
$\eta_q$ mesons.}
\begin{tabular}{l | c c c c c c c | c c c }
 & \multicolumn{7}{c|}{$A^T_\phi \phantom{ap}$} & \multicolumn{3}{c}{$A^T_{\phi\phi'}$ \phantom{sp}} \\
    & $\quad \pi \quad$ & $\quad K \quad $ & $\quad \eta_s \quad $ 
    & $ \quad ju \quad$ & $ \quad ru \quad$ 
    & $\quad js \quad$  & $\quad rs \quad$ 
    & $\quad \eta_u \eta_u \quad $ & $\quad \eta_u \eta_s\quad $   & $\quad \eta_s \eta_s\quad$ \\
\hline
$\D$       &  $\frac{2}{3}$ & $0$  & $0$  
           &  $\frac{2}{3}$ & $\frac{1}{3}$  
           &  $0$ & $0$
           &  $1$ & $0$ & $0$ \\

$\Sigma^*$ &  $\frac{2}{9}$ & $\frac{4}{9}$  & $0$  
           &  $\frac{4}{9}$ & $\frac{2}{9}$  
           &  $\frac{2}{9}$ & $\frac{1}{9}$
           &  $\frac{4}{9}$ & $\frac{4}{9}$ & $\frac{1}{9}$ \\

$\Xi^*$    &  $0$ & $\frac{4}{9}$  & $\frac{2}{9}$  
           &  $\frac{2}{9}$ & $\frac{1}{9}$  
           &  $\frac{4}{9}$ & $\frac{2}{9}$
           &  $\frac{1}{9}$ & $\frac{4}{9}$ & $\frac{4}{9}$ \\

$\Omega^-$ &  $0$ & $0$  & $\frac{2}{3}$  
           &  $0$ & $0$  
           &  $\frac{2}{3}$ & $\frac{1}{3}$
           &  $0$ & $0$ & $1$ \\
 
\end{tabular}
\label{t:PQQCD-A}
\end{table}

\begin{table}
\caption{The coefficients $B^T_\phi$ and $B^T_{\phi\phi'}$ in \PQCPT. Coefficients are
listed for the decuplet states $T$, and for $B^T_\phi$ are grouped into contributions from loop mesons
with mass $m_\phi$, while for $B^T_{\phi\phi'}$ are grouped into contributions from pairs of quark-basis 
$\eta_q$ mesons.}
\begin{tabular}{l | c c c c c c c | c c c }
 & \multicolumn{7}{c|}{$B^T_\phi \phantom{ap}$} & \multicolumn{3}{c}{$B^T_{\phi\phi'}$ \phantom{sp}} \\
    & $\quad \pi \quad$ & $\quad K \quad $ & $\quad \eta_s \quad $ 
    & $ \quad ju \quad$ & $ \quad ru \quad$ 
    & $\quad js \quad$  & $\quad rs \quad$ 
    & $\quad \eta_u \eta_u \quad $ & $\quad \eta_u \eta_s\quad $   & $\quad \eta_s \eta_s\quad$ \\
\hline
$\D$       &  $-\frac{2}{3}$ & $0$  & $0$  
           &  $\frac{4}{3}$ & $\frac{2}{3}$  
           &  $0$ & $0$
           &  $0$ & $0$ & $0$ \\

$\Sigma^*$ &  $-\frac{2}{9}$ & $-\frac{4}{9}$  & $0$  
           &  $\frac{8}{9}$ & $\frac{4}{9}$  
           &  $\frac{4}{9}$ & $\frac{2}{9}$
           &  $\frac{2}{9}$ & $-\frac{4}{9}$ & $\frac{2}{9}$ \\

$\Xi^*$    &  $0$ & $-\frac{4}{9}$  & $-\frac{2}{9}$  
           &  $\frac{4}{9}$ & $\frac{2}{9}$  
           &  $\frac{8}{9}$ & $\frac{4}{9}$
           &  $\frac{2}{9}$ & $-\frac{4}{9}$ & $\frac{2}{9}$ \\

$\Omega^-$ &  $0$ & $0$  & $-\frac{2}{3}$  
           &  $0$ & $0$  
           &  $\frac{4}{3}$ & $\frac{2}{3}$
           &  $0$ & $0$ & $0$ \\
 
\end{tabular}
\label{t:PQQCD-B}
\end{table}

To leading order in \PQCPT\ the decuplet masses are
\begin{equation}
M^{(1)}_T = \frac{2}{3} \gamma_M \, m_T  - 2 \, \ol \sigma_M \, \str (m_Q),
\label{eq:MPQLO}
\end{equation}
where the coefficients $m_T$ appear in Table \ref{t:mT}.
At next-to-leading order, we have
\begin{eqnarray}
M^{(3/2)}_T &=& -\frac{5 \cH^2}{72 \pi f^2} 
\left[ 
\sum_\phi A_\phi^T \, m_\phi^3 
+ 
\sum_{\phi\phi'} A_{\phi\phi'}^T \, \mathcal{M}^3 (m_\phi, m_{\phi'}) 
\right]
\notag \\ 
&-& \frac{\cC^2}{(4 \pi f)^2} 
\left[ \sum_\phi  B_\phi^T  \, \cF (m_\phi,-\D,\mu)
+
\sum_{\phi\phi'} B^T_{\phi\phi'} \, \cF (m_\phi, m_{\phi'}, -\D, \mu) \right]
,\end{eqnarray}

where $\mathcal{F}(m,\D,\mu)$ is given by Eq.~\eqref{eqn:F}. 
Additionally we have employed the abbreviations
\begin{align}
\mathcal{M}^n (m_\phi, m_{\phi'})  &=  \cH_{\phi\phi'}(m^n_\phi, m^n_{\phi'},m^n_X), \\
\cF (m_\phi, m_{\phi'}, \d, \mu) &=  
\cH_{\phi \phi'}[\cF(m_\phi,\d,\mu),\cF(m_{\phi'},\d,\mu), \cF(m_X,\d,\mu)]
\label{eq:MPQNLO}
,\end{align}
for contributions arising from the hairpin diagrams depicted in 
Figure \ref{fig:PQNLO}.
The sums involving these two functions are over pairs of flavor neutral states in the quark basis, e.g., above $\phi\phi'$ runs over
$\eta_u \eta_u$, $\eta_u \eta_s$, and $\eta_s \eta_s$. In this way there is no double counting.
The coefficients $A^T_\phi$ and $A^T_{\phi\phi'}$ appear in Table \ref{t:PQQCD-A} and $B^T_\phi$ and $B_{\phi\phi'}^T$
appear in Table \ref{t:PQQCD-B}.

At next-to-next-to leading order, we have the contribution to the decuplet mass
{\small
\begin{eqnarray}
M_T^{(2)} &=& 
(Z - 1) M_T^{(1)} 
\notag \\
&& + 
\frac{1}{4 \pi f} 
\left\{
\frac{1}{3} t_1^M \, (m^2)_T 
+ 
\frac{1}{3} t_2^M \, (m m')_T 
+ 
t_3^M \, \str(m_Q^2) 
+ 
\frac{1}{3} t_4^M \,  m_T  \, \str (m_Q)
+
t_5^M \, [\str (m_Q)]^2
\right\}
\notag \\
&& -
\frac{2 \,\gamma_M}{(4 \pi f)^2} 
\left[
\sum_\phi C_\phi^T \, \cL(m_\phi,\mu) 
+ 
\sum_{\phi\phi'} C_{\phi\phi'}^T \, \cL(m_\phi,m_{\phi'},\mu)
\right]
\notag \\
&& +
\frac{2 \, \ol \sigma_M}{(4 \pi f)^2} 
\left[
\sum_\phi \ol C_\phi \, \cL(m_\phi,\mu)
+
\sum_{\phi\phi'} \ol C_{\phi\phi'} \, \cL(m_\phi,m_{\phi'},\mu)
\right]
\notag \\
&& + 
\frac{1}{(4 \pi f )^3} 
\sum_\phi 
\left( 
t^A_1 \,  D_\phi^T + t^A_2 \, E_\phi^T + t^A_3 \, \ol D_\phi 
\right)
\ol \cL (m_\phi, \mu)
\notag \\
&& + 
\frac{1}{(4 \pi f )^3} 
\sum_{\phi \phi'} 
\left( 
t^A_1 \,  D_{\phi\phi'}^T + t^A_2 \, E_{\phi\phi'}^T + t^A_3 \, \ol D_{\phi\phi'} 
\right)
\ol \cL (m_\phi,m_{\phi'}, \mu)
\notag \\
&& + 
\frac{1}{4 (4 \pi f )^3} 
\sum_\phi 
\left[ 
(t^{\tilde{A}}_1  + t_1^{vA} ) D_\phi^T 
+ 
(t^{\tilde{A}}_2 + t_2^{vA} ) E_\phi^T 
+ 
(t^{\tilde{A}}_3  + t_3^{vA}) \ol D_\phi 
\right]
\left[
\ol \cL (m_\phi, \mu)
- 
\frac{1}{2}m_\phi^4 
\right]
\notag \\
&& + 
\frac{1}{4 (4 \pi f )^3} 
\sum_{\phi \phi'} 
\left[
(t^{\tilde{A}}_1  + t_1^{vA} ) D_{\phi\phi'}^T 
+ 
(t^{\tilde{A}}_2  + t_2^{vA} ) E_{\phi\phi'}^T 
+ 
(t^{\tilde{A}}_3  + t_3^{vA} ) \ol D_{\phi\phi'} 
\right]
\notag \\
&& \phantom{bigspacethattakesspace} \times
\left[
\ol \cL \left( m_\phi, m_{\phi'}, \mu \right)
-
\frac{1}{2} \mathcal{M}^4 (m_\phi, m_{\phi'})
\right]
\notag \\
&& -
\frac{5}{8} \frac{\cH^2}{(4\pi f)^2 M_B} 
\left\{ 
\sum_\phi A_\phi^T 
\left[ 
\ol \cL(m_\phi,\mu)
+ \frac{19}{10} m_\phi^4
\right]
+
\sum_{\phi\phi'} A_{\phi\phi'}^T 
\left[ 
\ol \cL \left(  m_\phi, m_{\phi'}, \mu \right)
+ \frac{19}{10} \mathcal{M}^4(m_\phi, m_{\phi'})
\right]
\right\}
\notag \\
&& - 
\frac{15}{16} \frac{\cC^2}{(4 \pi f)^2 M_B}  
\left\{
\sum_\phi B_\phi^T
\left[ 
\ol \cL(m_\phi,\mu)
- \frac{1}{10} m_\phi^4
\right]
+ 
\sum_{\phi\phi'} B_{\phi\phi'}^T 
\left[
\ol \cL \left( m_\phi, m_{\phi'}, \mu \right)
- \frac{1}{10} \mathcal{M}^4 (m_\phi, m_{\phi'})
\right]
\right\}
\notag \\
&& -
\frac{10 \cH^2 \ol \sigma_M \str(m_Q) }{3 (4 \pi f)^2} 
\left\{
\sum_\phi A_\phi^T 
\left[ 
\cL(m_\phi,\mu) + \frac{26}{15} m_\phi^2
\right]
+ 
\sum_{\phi \phi'} A_{\phi\phi'}^T 
\left[ 
\cL \left( m_\phi, m_{\phi'}, \mu \right)
+
\frac{26}{15} \mathcal{M}^2 (m_\phi, m_{\phi'}) 
\right]
\right\}
\notag \\
&&- 
\frac{3 \cC^2  \sigma_M \str(m_Q) }{ (4 \pi f)^2} 
\left[
\sum_\phi B_\phi^T  \, \cJ (m_\phi,-\D,\mu)
+
\sum_{\phi\phi'} B_{\phi\phi'}^T \, \cJ(m_\phi,m_{\phi'},-\D,\mu)
\right]
\notag \\
&& + 
\frac{10 \cH^2 \gamma_M}{9 (4 \pi f)^2}
\left\{
\sum_\phi F_\phi^T 
\left[ 
\cL(m_\phi,\mu) + \frac{26}{15} m_\phi^2
\right]
+
\sum_{\phi \phi'} F_{\phi\phi'}^T 
\left[ 
\cL \left( m_\phi, m_{\phi'}, \mu \right)
+ \frac{26}{15} \mathcal{M}^2(m_\phi, m_{\phi'})
\right]
\right\}
\notag \\
&&- 
\frac{3 \cC^2 }{2 (4 \pi f)^2} 
\left[
\sum_\phi G_\phi^T  \, \cJ (m_\phi,-\D,\mu)
+
\sum_{\phi\phi'} G_{\phi\phi'}^T \, \cJ(m_\phi,m_{\phi'},-\D,\mu)
\right]
\label{eq:MPQNNLO}
.\end{eqnarray}
}
The above expression is written quite compactly. It
involves the tree-level coefficients $m_T$,  $(m^2)_T$, and $(m m')_T$, 
which are listed in Table \ref{t:mT}, and the wavefunction renormalization $Z$, which 
we find to be
\begin{eqnarray}
Z - 1 &=& - \frac{5 \cH^2}{3 (4 \pi f)^2} 
\left\{
\sum_\phi A_\phi^T 
\left[ 
\cL(m_\phi,\mu)+ \frac{26}{15} m_\phi^2
\right]
+ 
\sum_{\phi\phi'} A_{\phi\phi'}^T 
\left[
\cL \left( m_\phi,m_{\phi'},\mu \right)  
+ \frac{26}{15} \mathcal{M}^2 (m_\phi, m_{\phi'})
\right]
\right\}
\notag \\
&&- \frac{3 \cC^2}{2 (4 \pi f)^2} 
\left[
\sum_\phi B_\phi^T  \cJ (m_\phi,-\D,\mu)
+
\sum_{\phi\phi'} B_{\phi\phi'}^T \cJ(m_\phi,m_{\phi'},-\D,\mu)
\right]
\label{eq:PQwfn}
.\end{eqnarray}
The new non-analytic functions arising from loop integrals in Eqs.~\eqref{eq:MPQNNLO} and
\eqref{eq:PQwfn} are defined to be
\begin{align}
\cL(m_\phi,m_{\phi'},\mu) &= \cH_{\phi\phi'}
[ \cL(m_\phi,\mu),\cL(m_{\phi'}, \mu), \cL(m_X,\mu)], \\   
\ol \cL(m_\phi,m_{\phi'},\mu) &= \cH_{\phi\phi'}
[ \ol \cL(m_\phi,\mu), \ol \cL(m_{\phi'}, \mu), \ol \cL(m_X,\mu)], \\ 
\cJ(m_\phi,m_{\phi'},\d,\mu) &= \cH_{\phi\phi'} 
[\cJ(m_\phi,\d,\mu),\cJ(m_{\phi'},\d,\mu),\cJ(m_X,\d,\mu) ]
\end{align}
and arise from the hairpin contributions shown in Figure \ref{fig:PQNNLO}. 
The various coefficients in the above sums are listed in Tables \ref{t:PQQCD-A}--\ref{t:PQQCD-G}. 
One may check that in the limit $m_j \to m_u$, $m_r \to m_s$ the \CPT\ results are obtained from 
the \PQCPT\ expressions. 

As in the \CPT\ case, the above expressions Eqs.~\eqref{eq:MPQLO}, \eqref{eq:MPQNLO}, and \eqref{eq:MPQNNLO} are quark mass
expressions, i.e.~, the meson masses in these expressions are merely replacements for the quark masses
via Eq.~\eqref{eq:mesonmass}. To utilize these expressions for chiral extrapolations of lattice QCD data, one must 
write the quark masses in terms of the lattice meson masses. To work to $\cO(m_q^2)$ means the LO contributions in Eq.~\eqref{eq:MPQLO}
must be written in meson masses out to NLO.

\begin{table}
\caption{The coefficients $C^T_\phi$, $C^T_{\phi\phi'}$, $\ol C_\phi$, and $\ol C_{\phi \phi'}$ in \PQCPT. The coefficients $\ol C_\phi$ and $\ol C_{\phi \phi'}$
are identical for all decuplet states. The remaining coefficients are
listed for the decuplet states $T$, and for $C^T_\phi$ are grouped into contributions from loop mesons
with mass $m_\phi$, while for $C^T_{\phi\phi'}$ are grouped into contributions from pairs of quark-basis 
$\eta_q$ mesons. If a particular meson or pair of flavor neutral mesons is not listed, then the value of the coefficient
is zero for all decuplet states. }
\begin{tabular}{l | c c c c | c c}
& \multicolumn{4}{c|}{$C^T_\phi \phantom{ap}$} & \multicolumn{2}{c}{$C^T_{\phi\phi'}$ \phantom{sp}} \\
      & $\quad ju \quad$ & $\quad ru \quad $ & $\quad js \quad $  & $\quad rs \quad$ & $\quad \eta_u \eta_u \quad$ & $\quad \eta_s \eta_s \quad$ \\
\hline
$\D$       
           &  $2( m_u + m_j)$ & $m_u + m_r$  & $0$ & $0$ 
	   &  $2 m_u$ & $0$ \\

$\Sigma^*$ 
           &  $\frac{4}{3}( m_u + m_j)$ & $\frac{2}{3}(m_u + m_r)$  & $\frac{2}{3}(m_s +  m_j)$ & $\frac{1}{3}(m_s + m_r)$
	   &  $\frac{4}{3} m_u$ & $\frac{2}{3} m_s$ \\

$\Xi^*$    
           &  $\frac{2}{3}(m_u + m_j)$ & $\frac{1}{3} (m_u + m_r)$  & $\frac{4}{3}( m_s +  m_j)$ & $\frac{2}{3} (m_s + m_r)$ 
	   &  $\frac{2}{3} m_u$ & $\frac{4}{3} m_s$ \\

$\Omega^-$ 
           &  $0$ & $0$  & $2 (m_s + m_j)$ & $m_s + m_r$ 
	   &  $0$ & $2 m_s$ \\
\multicolumn{6}{c}{}
\\
&  $\quad jj \quad$ & $\quad jr \quad$ & $\quad rr \quad$ &  & $\quad \eta_j \eta_j \quad$ & $\quad \eta_r \eta_r \quad$ \\  
\hline 
$\ol C_\phi$	   & $8 m_j$ & $4(m_j + m_r)$ & $2 m_r$ & $\quad \quad \quad \ol C_{\phi\phi'}$ & $4 m_j$ & $2 m_r$
\end{tabular}
\label{t:PQQCD-C}
\end{table}

\begin{table}
\caption{The coefficients $D^T_\phi$, $D^T_{\phi\phi'}$, $\ol D_\phi$, and $\ol D_{\phi \phi'}$ in \PQCPT. 
The coefficients $\ol D_\phi$ and $\ol D_{\phi \phi'}$
are identical for all decuplet states. The remaining coefficients are
listed for the decuplet states $T$, and for $D^T_\phi$ are grouped into contributions from loop mesons
with mass $m_\phi$, while for $D^T_{\phi\phi'}$ are grouped into contributions from pairs of quark-basis 
$\eta_q$ mesons. If a particular meson or pair of flavor neutral mesons is not listed, then the value of the coefficient
is zero for all decuplet states.}
\begin{tabular}{l | c c c c | c c}
& \multicolumn{4}{c|}{$D^T_\phi \phantom{ap}$} & \multicolumn{2}{c}{$D^T_{\phi\phi'}$ \phantom{sp}} \\
      & $\quad ju \quad$ & $\quad ru \quad $ & $\quad js \quad $  & $\quad rs \quad$ & $\quad \eta_u \eta_u \quad$ & $\quad \eta_s \eta_s \quad$ \\
\hline
$\D$       
           &  $2$ & $1$  & $0$ & $0$ 
	   &  $1$ & $0$ \\

$\Sigma^*$ 
           &  $\frac{4}{3}$ & $\frac{2}{3}$  & $\frac{2}{3}$ & $\frac{1}{3}$
	   &  $\frac{2}{3}$ & $\frac{1}{3}$ \\

$\Xi^*$    
           &  $\frac{2}{3}$ & $\frac{1}{3}$  & $\frac{4}{3}$ & $\frac{2}{3}$ 
	   &  $\frac{1}{3}$ & $\frac{2}{3}$ \\

$\Omega^-$ 
           &  $0$ & $0$  & $2$ & $1$ 
	   &  $0$ & $1$ \\
\multicolumn{6}{c}{}
\\
&  $\quad jj \quad$ & $\quad jr \quad$ & $\quad rr \quad$ &  & $\quad \eta_j \eta_j \quad$ & $\quad \eta_r \eta_r \quad$ \\  
\hline 
$\ol D_\phi$	   & $4$ & $4$ & $1$ & $\quad \quad \quad \ol D_{\phi\phi'}$ & $2$ & $1$
\end{tabular}
\label{t:PQQCD-D}
\end{table}

\begin{table}
\caption{The coefficients $E^T_\phi$ and $E^T_{\phi\phi'}$ in \PQCPT. Coefficients are
listed for the decuplet states $T$, and for $E^T_\phi$ are grouped into contributions from loop mesons
with mass $m_\phi$, while for $E^T_{\phi\phi'}$ are grouped into contributions from pairs of quark-basis 
$\eta_q$ mesons.}
\begin{tabular}{l | c c c | c  c c}
& \multicolumn{3}{c|}{$E^T_\phi \phantom{ap}$} & \multicolumn{3}{c}{$E^T_{\phi\phi'}$ \phantom{sp}} \\
      & $\quad \pi \quad$ & $\quad K \quad $ & $\quad \eta_s \quad $  & $\quad \eta_u \eta_u \quad$ & $\quad \eta_{u} \eta_s \quad $& $\quad \eta_s \eta_s \quad$ \\
\hline
$\D$       
           &  $1$ & $0$  & $0$ 
	   & $1$  & $0$  & $0$ \\

$\Sigma^*$ 
           &  $\frac{1}{3}$ & $\frac{2}{3}$  & $0$ 
           &  $\frac{1}{3}$ & $\frac{2}{3}$  & $0$ \\

$\Xi^*$    
           &  $0$ & $\frac{2}{3}$  & $\frac{1}{3}$ 
	   &  $0$ & $\frac{2}{3}$  & $\frac{1}{3}$ \\

$\Omega^-$ 
           &  $0$ & $0$  & $1$ 
	   &  $0$ & $0$ & $1$ 
\end{tabular}
\label{t:PQQCD-E}
\end{table}

\begin{table}
\caption{The coefficients  $F^T_\phi$ and $F^T_{\phi\phi'}$ in \PQCPT. Coefficients are
listed for the decuplet states $T$, and for $F^T_\phi$ are grouped into contributions from loop mesons
with mass $m_\phi$, while for $F^T_{\phi\phi'}$ are grouped into contributions from pairs of quark-basis 
$\eta_q$ mesons.}
\begin{tabular}{l | c c c c }
& \multicolumn{4}{c}{$F^T_\phi \phantom{ap}$} \\
      & $\quad \pi \quad$ & $\quad K \quad $ & $\quad \eta_s \quad $ & \\  
\hline
$\D$       
           &  $2 m_u$ & $0$  & $0$ &\\

$\Sigma^*$ 
	   &  $\frac{2}{9}(2 m_u + m_s)$ & $\frac{4}{9} ( 2 m_u + m_s)$  & $0$ &\\            

$\Xi^*$    
           &  $0$ & $\frac{4}{9} ( m_u + 2 m_s)$  & $\frac{2}{9}(m_u + 2 m_s)$ &\\ 

$\Omega^-$ 
	   &  $0$ & $0$  & $2 m_s$ & \\
\multicolumn{5}{c}{} 
\\
 	& $\quad ju \quad$ & $\quad ru \quad$ & $\quad js \quad$ & $\quad rs \quad$ \\
\hline
$\D$    &  $\frac{2}{3} (2 m_u + m_j)$ & $\frac{1}{3}(2 m_u + m_r)$  & $0$ & $0$ \\
$\Sigma^*$  &  $\frac{4}{9}(m_u + m_s + m_j)$ & $\frac{2}{9}(m_u + m_s + m_r)$  & $\frac{2}{9} (2 m_u + m_j)$ & $\frac{1}{9} (2 m_u + m_r)$ \\
$\Xi^*$    &  $\frac{2}{9}(2 m_s + m_j)$ & $\frac{1}{9}(2 m_s + m_r)$  & $\frac{4}{9}(m_u + m_s + m_j)$ & $\frac{2}{9}(m_u + m_s + m_r)$ \\
$\Omega^-$   &  $0$ & $0$  & $\frac{2}{3}(2 m_s + m_j)$ & $\frac{1}{3} (2 m_s + m_r)$ \\
\multicolumn{5}{c}{} 
\\
& \multicolumn{4}{c}{$F^T_{\phi\phi'}$ \phantom{sp}} 
\\
    	& $\quad \eta_u \eta_u \quad$ & $\quad \eta_{u} \eta_s \quad $& $\quad \eta_s \eta_s \quad$ &\\
\hline
$\D$   &  $3 m_u$ & $0$  & $0$ &\\
$\Sigma^*$  &  $\frac{4}{9}(2 m_u + m_s)$   & $\frac{4}{9}(2 m_u + m_s)$  & $\frac{1}{9}(2 m_u + m_s)$ & \\
$\Xi^*$ &  $\frac{1}{9}(m_u + 2 m_s)$  & $\frac{4}{9}(m_u + 2 m_s)$  & $\frac{4}{9}(m_u + 2 m_s)$ & \\
$\Omega^-$  &  $0$ & $0$  & $3 m_s$ & 
\end{tabular}
\label{t:PQQCD-F}
\end{table}

\begingroup
\squeezetable
\begin{table}
\caption{The coefficients $G^T_\phi$ and $G^T_{\phi\phi'}$ in \PQCPT. Coefficients are
listed for the decuplet states $T$, and for $G^T_\phi$ are grouped into contributions from loop mesons
with mass $m_\phi$, while for $G^T_{\phi\phi'}$ are grouped into contributions from pairs of quark-basis 
$\eta_q$ mesons.}
\begin{tabular}{l | c c c c }
& \multicolumn{4}{c}{$G^T_\phi \phantom{ap}$} \\
      & $\quad \pi \quad$ & $\quad K \quad $ & $\quad \eta_s \quad $ & \\  
\hline
\hline

$\D$       
&  $- \frac{4}{3} m_u (\a_M + \b_M)$ 
&  $0$  
&  $0$ &\\

\hline

$\Sigma^*$
&  $-\frac{4}{27}[m_u(\a_M + 4 \b_M)$ 
&  $-\frac{4}{27}[m_u(5 \a_M + 2 \b_M)$  
&  $0$ &\\            

& $\qquad + m_s (2 \a_M - \b_M)]$
& $\qquad + m_s (\a_M + 4\b_M)]$
& $$ &\\

\hline
$\Xi^*$    
&  $0$ 
&  $-\frac{4}{27}[m_u(\a_M + 4\b_M)$  
&  $-\frac{4}{27}[m_u(2 \a_M - \b_M)$ &\\

& $$
& $\qquad + m_s (5 \a_M + 2 \b_M)]$
& $\qquad + m_s (\a_M + 4 \b_M)]$ &\\
 
\hline
$\Omega^-$
&  $0$ 
&  $0$  
&  $- \frac{4}{3} m_s (\a_M + \b_M)$ & \\

\multicolumn{5}{c}{} 
\\
 	& $\quad ju \quad$ & $\quad ru \quad$ & $\quad js \quad$ & $\quad rs \quad$ \\
\hline
\hline

$\D$    
&  $\frac{4}{9}[m_u (5 \a_M + 2 \b_M) $ 
&  $\frac{2}{9}[m_u (5 \a_M + 2 \b_M)$  
&  $0$ 
&  $0$ \\

& $\qquad + m_j ( \a_M + 4 \b_M)]$
& $\qquad + m_r ( \a_M + 4 \b_M)]$
& $$
& $$ \\

\hline

$\Sigma^*$  
&  $\frac{4}{27} [ m_u (5 \a_M + 2 \b_M)$ 
&  $\frac{2}{27} [ m_u (5 \a_M + 2 \b_M)$  
&  $\frac{4}{27} [ m_u (5 \a_M + 2 \b_M)$ 
&  $\frac{2}{27} [ m_u (5 \a_M + 2 \b_M)$ \\

& $\qquad + m_s (5 \a_M + 2 \b_M)$
& $\qquad + m_s (5 \a_M + 2 \b_M)$
& $\qquad + m_j ( \a_M + 4 \b_M)]$
& $\qquad + m_r ( \a_M + 4 \b_M)]$ \\

& $\qquad + 2 m_j ( \a_M + 4 \b_M)]$
& $\qquad + 2 m_r ( \a_M + 4 \b_M)]$
& $$
& $$ \\

\hline

$\Xi^*$    
&  $\frac{4}{27} [ m_s (5 \a_M + 2 \b_M)$ 
&  $\frac{2}{27} [ m_s (5 \a_M + 2 \b_M)$  
&  $\frac{4}{27} [ m_u (5 \a_M + 2 \b_M)$ 
&  $\frac{2}{27} [ m_u (5 \a_M + 2 \b_M)$ \\

& $\qquad + m_j ( \a_M + 4 \b_M)]$
& $\qquad + m_r ( \a_M + 4 \b_M)]$
& $\qquad + m_s (5 \a_M + 2 \b_M)$
& $\qquad + m_s (5 \a_M + 2 \b_M)$ \\

& $$
& $$
& $\qquad + 2 m_j ( \a_M + 4 \b_M)]$
& $\qquad + 2 m_r ( \a_M + 4 \b_M)]$ \\

\hline

$\Omega^-$   
&  $0$ 
&  $0$  
&  $\frac{4}{9}[m_s (5 \a_M + 2 \b_M)$ 
&  $\frac{2}{9}[m_s (5 \a_M + 2 \b_M)$ \\

& $$
& $$
& $\qquad + m_j ( \a_M + 4 \b_M)]$
& $\qquad + m_r ( \a_M + 4 \b_M)]$ \\
\multicolumn{5}{c}{} 
\\
& \multicolumn{4}{c}{$G^T_{\phi\phi'}$ \phantom{sp}} 
\\
    	& $\quad \eta_u \eta_u \quad$ & $\quad \eta_{u} \eta_s \quad $& $\quad \eta_s \eta_s \quad$ &\\
\hline
\hline
$\D$   
&  $0$ 
&  $0$  
&  $0$ &\\

\hline
$\Sigma^*$  
&  $\frac{2}{27}[m_u (5 \a_M + 2 \b_M)$   
&  $-\frac{4}{27}[ m_u (5 \a_M + 2 \b_M)$  
&  $\frac{2}{27} [m_u(5 \a_M + 2 \b_M)$ & \\

& $\qquad + m_s (\a_M + 4 \b_M)]$ 
& $\qquad + m_s (\a_M + 4 \b_M)]$
& $\qquad + m_s (\a_M + 4 \b_M)]$ & \\
\hline

$\Xi^*$ 
&  $\frac{2}{27}[m_u (\a_M + 4 \b_M)$  
&  $-\frac{4}{27}[ m_u (\a_M + 4 \b_M)$  
&  $\frac{2}{27}[m_u (\a_M + 4 \b_M)$ & \\

& $\qquad + m_s (5 \a_M + 2 \b_M)]$
& $\qquad + m_s (5 \a_M + 2 \b_M)]$
& $\qquad + m_s (5 \a_M + 2 \b_M)]$ & \\

\hline

$\Omega^-$  
&  $0$ 
&  $0$  
&  $0$ & 
\end{tabular}
\label{t:PQQCD-G}
\end{table}
\endgroup

\section{Summary} \label{summy}

We have calculated the masses of the decuplet baryons
in the isospin limit of three-flavor \CPT\ and
have also derived the decuplet masses in the analogous
partially quenched theory.  We have kept all relevant terms
to $\cO(m_q^2)$, including terms fixed by reparameterization 
invariance that ensure the Lorentz invariance of the heavy
baryon effective theory. Knowledge of the low-energy
behavior of QCD and PQQCD is crucial to properly extrapolate
lattice calculations from the light quark masses used on the 
lattice to those in nature.

Working to $\cO(m_q^2)$
in the chiral expansion forces the introduction of a large
number of low-energy constants. For any predictive power, 
the low-energy constants must be fit from experiment
or lattice results. Lattice calculations will eventually 
allow first principles determination of these constants
and thus predictions from QCD. In the foreseeable future, 
partially quenched simulations will enable these rigorous 
predictions. Our \PQCPT\ results for decuplet
baryon masses are required for the proper extrapolation of PQQCD lattice
data and hence for physical predictions from QCD. 
As the decuplet baryons are resonances, a short procedural comment 
is in order.  For large enough pion masses, 
the decuplet states are stable to strong decays and can be 
studied on the lattice, \emph{cf}. our expressions are all real valued for
these pion masses. Thus for 
$m_\pi \sim 300 \, \text{MeV}$,\footnote{%
This is a conservative estimate. Volume effects will modify
the decay threshold of the decuplet.
}
where one still trusts the chiral expansion, 
the decuplet properties can be calculated on the lattice.
One then uses the expressions we have derived to fit the low-energy constants. 
The expressions at next-to-next-to-leading order are necessary for
reducing uncertainty in the chiral extrapolation. Armed with the low-energy 
constants, one can then make predictions for the decuplet resonances, e.g.~%
their decay widths can be found via
\begin{eqnarray}
\Im \text{m} (M_T) 
&=& 
- \frac{\cC^2}{8 \pi f^2} 
(\D^2 - m_\pi^2)^{3/2} \, B^T_\pi
\\
&-& 
\frac{\cC^2}{4 \pi f^2} 
\D \sqrt{\D^2 - m_\pi^2}
\left\{
\left[
\gamma_M \, m_T  + 3 (\sigma_M - \ol \sigma_M ) \tr (m_Q) 
\right]
\, B^T_\pi
+ \frac{3}{2} \, G_\pi^T
\right\}
.\end{eqnarray}

To reiterate: the low-energy constants appearing in \CPT\
appear in \PQCPT\ and by fitting them using PQQCD lattice calculations
one can make QCD predictions. The decuplet baryon masses are no exception;
our \PQCPT\ results exhibit a smooth limit to \CPT\ as the sea quark masses vary. 
Furthermore, lattice simulations with unphysically stable decuplet states 
can be used in conjunction with \CPT\ and \PQCPT\  to predict properties of 
the physical resonances.

\acknowledgments
We thank Martin Savage for many useful discussions
and David Lin for comments on the manuscript.
This work was supported in part by the U.S. Department of
Energy under Grant No. DE-FG03-97ER4014.

\bibliography{hb}
\end{document}